\documentclass[11pt]{article}
\pdfoutput=1

\usepackage{mjheppub}
\usepackage[T1]{fontenc} 

\usepackage{longtable}
\usepackage[vcentermath]{youngtab}
\usepackage{array,multirow}
 \usepackage{multirow}
\usepackage{amssymb}
\usepackage{amsfonts}
\usepackage{amsbsy}
\usepackage{amsmath}
\usepackage{amsthm}
\usepackage{graphicx}
\usepackage{epstopdf}
\usepackage{color}
\usepackage{hyperref}
\usepackage{wasysym}

\makeatletter
\g@addto@macro\bfseries{\boldmath}
\def\NAT@sort{\z@}
\makeatother

\def \be {\begin{equation}}
\def \ee {\end{equation}}
\def \bsp {\begin{split}}
\def \esp {\end{split}}
\def \bea {\begin{eqnarray}}
\def \eea {\end{eqnarray}}

\def\Z{\mathbb{Z}}
\def\F{\mathbb{F}}

\def\P{\mathbb{P}}

\def\n3a{t}

\definecolor{Red}{rgb}{1.00, 0.00, 0.00}

\title{
Fibration structure\\  in toric hypersurface Calabi-Yau threefolds
}
\author{Yu-Chien Huang,}
\author{Washington Taylor}

\affiliation{Center for Theoretical Physics\\
Department of Physics\\
Massachusetts Institute of Technology\\
77 Massachusetts Avenue\\
Cambridge, MA 02139, USA}

\emailAdd{{\tt yc\_huang} {\rm at} {\tt mit.edu}}
\emailAdd{{\tt wati} {\rm at} {\tt mit.edu}}

\preprint{MIT-CTP-5132}

\abstract{
We find through a systematic analysis that all but 29,223 of the 473.8
million 4D reflexive polytopes found by Kreuzer and Skarke have a 2D
reflexive subpolytope.  Such a subpolytope is generally associated
with the presence of an elliptic or genus one fibration in the
corresponding birational equivalence class of Calabi-Yau threefolds.
This extends the growing body of evidence that most Calabi-Yau
threefolds have an elliptically fibered phase.}

\begin{document}
\maketitle

\flushbottom


\section{Introduction}
\label{sec:intro}

Calabi-Yau manifolds have been studied intensively by physicists and
mathematicians over the last several decades, since the first
realization that these geometric spaces can be used to construct
supersymmetric compactifications of string theory \cite{chsw}.
Despite much work, the question of whether the number of distinct
topological types of Calabi-Yau threefolds is finite or
infinite  \cite{Yau-66} is still unresolved.  The special class of Calabi-Yau
manifolds that have an elliptic or genus one fibration, on the other
hand, is better understood mathematically and it has been proven that
there are a finite number of distinct Calabi-Yau threefolds of this
type, up to birational equivalence \cite{Gross}.  In recent years,
elliptic Calabi-Yau manifolds have been studied in great detail by
physicists due to their role in the nonperturbative formulation of
string theory known as F-theory \cite{Vafa-F-theory, Morrison-Vafa-I,
  Morrison-Vafa-II}.  Using a combination of methods and insights from
mathematics and physics, and building on Grassi's results for minimal
models of the bases that support elliptic fibrations \cite{Grassi}, we
have a good global understanding of the connected space of elliptic
Calabi-Yau threefolds, the bounds on their Hodge numbers, and a
systematic approach to constructing such threefolds that can be used
to essentially enumerate all elliptic Calabi-Yau threefolds at large
$h^{2,1}$, though there are technical issues that make a complete
enumeration at smaller values of $h^{2,1}$ currently out of reach (see
e.g. \cite{KMT-2, toric, Hodge, Wang-WT, Johnson-WT, Johnson-WT-2}).

There is a growing body of evidence that
most known Calabi-Yau threefolds are in fact
elliptic or genus one fibered (at least up to birational equivalence).  Some of the largest known sets of
Calabi-Yau threefolds are the 7,890 complete intersection Calabi-Yau (CICY)
threefolds \cite{cdls}, more generalized complete intersection
Calabi-Yaus
\cite{Anderson-aggl},
and the  Calabi-Yau threefolds constructed as
hypersurfaces in toric varieties associated with the 473.8 million
reflexive 4D polytopes \cite{KS-classification}.  It was
shown in \cite{aggl-2, aggl-3} that 99.3\% of the CICY threefolds have
an ``obvious'' genus one or elliptic fibration, and that many of these
threefolds admit many such fibrations; similar results hold for
discrete quotients of the CICY threefolds \cite{agh-non-simply}.
It was shown in \cite{Candelas-cs} that many polytopes in the
Kreuzer-Skarke (KS) database
\cite{database} have a structure compatible with a K3 fibration.
A systematic construction of
elliptic CY threefolds at large Hodge numbers over toric base surfaces
\cite{Huang-Taylor-long}
showed that all Hodge number pairs in the  KS database
with $h^{1,1} \geq 240$ or $h^{2,1}
\geq 240$ are associated with such elliptic Calabi-Yau threefolds.
A
systematic direct study of the fibration structure of the polytopes in the KS database was initiated in
\cite{Huang-Taylor}; in that paper we found that all polytopes
associated with Calabi-Yau threefolds having $h^{1,1} \geq 150$ or
$h^{2,1} \geq 150$ have a reflexive 2D subpolytope, indicating a
structure compatible with the presence of an elliptic or genus one
fibration for the associated CY threefolds.  In that paper we also
found empirically  that at small $h^{1,1}$, the fraction of 4D
polytopes lacking such a reflexive 2D subpolytope drops roughly as
$2^{- h^{1,1}}$, and we gave some analytic arguments for why these
results may naturally be expected.

In this paper we complete the program begun in \cite{Huang-Taylor},
and we report on the results of a complete analysis of all 473.8
million reflexive 4D polytopes for reflexive 2D subpolytopes.  
The upshot of the analysis is that  the 
Calabi-Yau threefolds that correspond to the toric hypersurface
construction seem to be dominated by those that are elliptic or genus
one fibered in some phase, so that
in fact most known Calabi-Yau threefolds
are birationally equivalent to one with an elliptic or genus one
fibration.  
Section
\ref{sec:subpolytopes} contains a description of the methodology used
and a brief discussion of how the existence of a
reflexive 2D subpolytope is related to the existence of an elliptic or genus one
fibration of the associated Calabi-Yau threefolds.
Section~\ref{sec:results} describes the results of the subpolytope
analysis, and Section~\ref{sec:discussion} contains a summary of the
results and discussion.

\section{Reflexive subpolytopes and fibrations}
\label{sec:subpolytopes}

The basic question we consider here is which reflexive 4D polytopes
have 2D subpolytopes that pass through the origin and hence act as
fibers.  From the work of Batyrev \cite{Batyrev}, we know that a
hypersurface in a toric variety associated with a reflexive 4D polytope
generically gives a smooth Calabi-Yau threefold; thus, a reflexive 2D
subpolytope $\nabla_2$ of a reflexive 4D polytope $\nabla$ suggests the presence of a
genus one or elliptic fibration for a corresponding Calabi-Yau
threefold \cite{Kreuzer-Skarke-fibers}.

The problem of identifying 2D subpolytopes from the combinatorial data
of a 4D polytope is described and discussed in some detail in
\cite{Rohsiepe, Braun, Huang-Taylor}.  We use here the notation and conventions
of \cite{Huang-Taylor-long, Huang-Taylor}, to which the reader is referred for further
background and references.

\subsection{Methodology}
\label{sec:methodology}

The algorithm that we have used to identify the existence of a 2D
reflexive subpolytope is a streamlined version of the algorithms
discussed in \cite{Rohsiepe, Braun, Huang-Taylor}.  The basic idea is to
determine whether any pair of rays in $\nabla \cap \Z^4$ generate a
2D sublattice of $\Z^4$ that intersects $\nabla$ in a set of points
that form a 2D polytope containing the origin as an interior point.

We use as starting data the set of lattice points $L = \nabla \cap\Z^4 \subset \nabla$
and the set of vertices $V$ of the dual polytope 
$\Delta = \nabla^*=\{w: \langle w,p\rangle\geq-1, \forall p\in \nabla\}$.  For
each lattice point $p \in L$, we compute
\begin{equation}
s(p) = {\rm max}_{v \in V} \langle p, v \rangle \,.
\end{equation}
We then make a list of all the points in $L$ with $s (p) \in\{1, 2,
3\}$
\begin{equation}
 S_i =\{p \in L: s (p) = i\} \,.
\label{eq:si}
\end{equation}
The purpose of this set of lists is to reduce the number of
computations needed to check for the  presence of the 2D subpolytope; the data in $S_1,
S_2, S_3$ is all that is needed to check whether such a subpolytope
exists.  This reduces the computational cost of the analysis
substantially when the number of points in $L$ is large.

There are 16 distinct 2D reflexive polytopes, discussed in the physics
context in \cite{Bouchard-Skarke, Braun, BGK-geometric, Klevers-16}.
For each of these reflexive polytopes, it is straightforward to check
whether one of the following conditions holds:
\begin{enumerate}
\item[I] $\exists x, y \in S_1: x \neq y$ and $- x, - y \in S_1$
\item[II]  $\exists x, y \in S_2:  - (x + y) \in S_1$ or $ - (x + y) \in S_2$
\item[III]  $\exists x, y \in S_3: -(x + y)/2 \in S_1$.
\end{enumerate}
In particular, with the standard ordering of the 16 fiber polytopes (used in e.g.\ \cite{Braun}
and \cite{Huang-Taylor}), condition I holds for fibers 2, 5, 7, 8, 9,
11, 12, 13, 14, 15, 16; of the fibers where condition I fails,
condition II holds for fibers 1, 3, 6, 10; and
condition III holds for the sole remaining case, fiber 4.  By simply checking these conditions
for the $S_i$ for each polytope, we can ascertain whether or not there is a
reflexive 2D subpolytope $\nabla_2$ of $\nabla$.

Note that once the existence of a 2D subpolytope $\nabla_2$
is determined, it is
straightforward to identify the full set of lattice points from
$\nabla$ that are in $\nabla_2$ as those that lie in the plane spanned by the pair of points
$x, y$ that satisfy one of the above 3 conditions.
The specific fiber type of $\nabla_2$ is then uniquely determined by
the numbers of points $n_i = | \nabla_2 \cap S_i |$ for $i = 1,
\ldots, 5$.  For a determination of fiber types it is therefore
helpful to initially determine the sets (\ref{eq:si}) for $i = 1,
\ldots, 5$, and then we only need to compute which of these lattice
points lie in the plane spanned by $x, y$.   The values of $n_i$
associated with each of the 16 fiber types are listed in
Table~\ref{t:types}. This information is also contained in Appendix A
of \cite{HT-mirror}; the numbers associated with the lattice points in
each reflexive 2D polytope there are $n_i +1$ in this notation.
While it is thus straightforward in any given case to determine the
fiber type, we have not collected this information in the full
database study carried out here, and in most of this paper we focus
simply on the binary question for each 4D reflexive polytope of
whether there is a 2D reflexive subpolytope fiber.  A more detailed
analysis of the fiber types at large and small Hodge numbers was
carried out in \cite{Huang-Taylor}.

\begin{table}
  \begin{center}
\begin{tabular}{ | c | c | } 
  \hline
  Fiber number &  $(n_1, \ldots, n_5)$\\
  \hline
  1 & (0, 3, 0, 0, 0)\\
  2 & (4, 0, 0, 0, 0)\\
  3 & (2, 2, 0, 0, 0)\\
  4 & (2, 0, 2, 0, 0)\\
  5 & (4, 1, 0, 0, 0)\\
  6 & (2, 2, 1, 0, 0)\\
  7 & (6, 0, 0, 0, 0)\\
  8 & (4, 1, 1, 0, 0)\\
  9 & (4, 2, 0, 0, 0)\\
  10 & (2, 2, 1, 0, 1)\\
  11 & (4, 2, 1, 0, 0)\\
  12 & (6, 1, 0, 0, 0)\\
  13 & (4, 2, 2, 0, 0)\\
  14 & (6, 2, 0, 0, 0)\\
  15 & (8, 0, 0, 0, 0)\\
  16 & (6, 3, 0, 0, 0)\\
  \hline
\end{tabular}
\end{center}
\caption[x]{\footnotesize Distinguishing values of $n_i
 = | \nabla_2 \cap S_i |$ for the 16 fibers}
\label{t:types}
\end{table}

\subsection{Implementation}
\label{sec:implementation}

To carry out the calculation over all 473.8 million reflexive 4D
polytopes, we first organized the information in the KS database
\cite{database} in files indexed by the Hodge numbers of the
associated Calabi-Yau threefolds.  We then used the {\it palp}
software package to compute the set of lattice points $L \subset
\nabla$ and dual vertices $V \subset \Delta$ for each polytope.
The core algorithm described in the previous subsection to identify 2D
reflexive subpolytopes was implemented in the language {\it Julia},
which combines ease of programming with efficiency through
just-in-time compiling.  Checking all 473.8 million cases took roughly
4 days on a single core of a Linux laptop.  The Julia code is
available at \cite{data}.

\subsection{Fibrations of polytopes versus toric varieties}
\label{sec:comparison}

While identifying a reflexive 2D subpolytope of the 4D polytope
$\nabla$ suggests the presence of an elliptic or genus one fibration
for the associated toric varieties, there are some subtleties in
making this connection precise.\footnote{Thanks to Antonella Grassi
  for emphasizing the importance of these issues.}  While some
analyses of fibration structures have focused on the structure of the
polytope alone -- sometimes in the language of ``tops'' \cite{tops,
  Bouchard-Skarke} -- as mentioned in \cite{Kreuzer-Skarke-fibers} and
discussed in more detail in \cite{Rohsiepe} not every triangulation of
a 4D polytope $\nabla$ with a 2D reflexive subpolytope $\nabla_2$
leads to a toric morphism that maps the subpolytope $\nabla_2$ to the
origin of the toric fan for the natural corresponding 2D base.  A
condition stated in \cite{Kreuzer-Skarke-fibers} is that the base $B$
of such a fibration should be identified by constructing the 2D toric
variety from the set of primitive rays in the image of $\nabla$ under
the projection that takes $\nabla_2 \rightarrow 0$.  Indeed, in many
cases, such as the ``standard stacking'' polytopes corresponding to
many generic and (Tate) tuned Weierstrass models over a given base,
one can use the structure of the polytopes and tops to determine the
base and additional tuned Kodaira singularity types to directly
construct a Weierstrass model for an elliptic fibration over the given
base, thus identifying a Calabi-Yau threefold that is elliptically
fibered and has the requisite Hodge numbers associated with the
polytope, circumventing the explicit construction of fans compatible
with a toric morphism from $\nabla$ to $\nabla_2$
\cite{Huang-Taylor-long}.  Particularly for more complicated
fibrations with general fibers and twists, however, there is no
systematic methodology for implementing such a direct construction;
and in any case, it is desirable to know in general which
triangulations of $\nabla$ are compatible with the fibration
structure, and whether in fact such triangulations always exist.

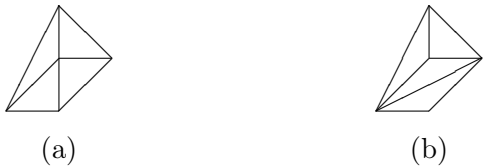
\begin{figure}
\begin{center}
\begin{picture}(200,60)(- 100,- 30)
\put(-90,-20){\line(1,0){20}}
\put(-90,-20){\line(1,1){20}}
\put(-90,-20){\line(1,2){20}}
\put(-70,0){\line(1,0){20}}
\put(-70,-20){\line(0,1){40}}
\put(-70,-20){\line(1,1){20}}
\put(-70,20){\line(1, -1){20}}
\put(-70, -35){\makebox(0,0){(a)}}
\put(50,-20){\line(1,0){20}}
\put(50,-20){\line(1,1){20}}
\put(50,-20){\line(1,2){20}}
\put(50,-20){\line(2,1){40}}
\put(70,0){\line(1,0){20}}
\put(70,0){\line(0,1){20}}
\put(70,-20){\line(1,1){20}}
\put(70,20){\line(1, -1){20}}
\put(70, -35){\makebox(0,0){(b)}}
\end{picture}
\end{center}
\caption[x]{\footnotesize Two triangulations of the 2D face
  corresponding to the base resulting from distinct Calabi-Yau phase
  triangulations of the 4D polytope associated with the generic
  elliptic fibration over the Hirzebruch surface $\F_1$; the first
  triangulation gives a toric variety with a toric morphism to the
  base $\F_1$, while the second triangulation gives a variety with a
  non-flat toric morphism to $\P^2$.}
\label{f:triangulations}
\end{figure}

Some of these questions regarding triangulations will be addressed in
detail elsewhere \cite{Per-forthcoming, cj-forthcoming}; here we
simply summarize a few key points.  First, for a given polytope, in
addition to triangulations compatible with a toric morphism to the
base associated with the full set of rays in the image of the
projection, there can also be triangulations compatible only with a
toric morphism to a base with smaller $h^{1,1}$, which are associated
with non-flat elliptic fibrations.\footnote{Non-flat elliptic
  fibrations were encountered and described in this toric context in
  \cite{Rohsiepe, Braun}; non-flat fibrations were and analyzed in more general
  physics and F-theory contexts in \cite{Candelas-nf,
    Lawrie-Schafer-Nameki, Borchmann:2013jwa, Cvetic:2013uta,
aggl-2, dor-nf, Huang-Taylor-long,
Achmed-Zade:2018idx,    Tian-Wang, alm-nf, Apruzzi:2019opn}.} An example is
given by the ``standard stacking'' polytope associated with the
generic elliptic fibration over the base $\F_1$ with the fiber $\P^{2,
  3, 1}$.  The vertices of this polytope are $(0, 0; 1, 0), (0, 0; 0,
1), (u^{(B)}_i; - 2, -3)$, where $u^{(B)}_i = (0, 1), (1, 0), (0, -1), (-1, -1)$
are the vertices of the polytope associated with the base $\F_1$.  The
resulting polytope $\nabla$ has triangulations associated with two
different Calabi-Yau geometries related by a flop.  The part of these
triangulations associated with the 2D face $B \cong\F_1$ of $\nabla$
is shown in Figure~\ref{f:triangulations}.  The first triangulation
leads to a toric morphism associated with an elliptic fibration over
$\F_1$.  The second triangulation, on the other hand, corresponds to a
non-flat toric morphism giving a non-flat elliptic fibration in which
the divisor associated with $v_3$ is contracted to a point in the
toric base $\P^2$.  For bases with larger 2D polytopes, associated
with Calabi-Yau threefolds with larger $h^{1,1}$, we expect many more
non-flat elliptic and genus one fibrations of this type to arise from
triangulations of this kind.  In other cases, triangulations are
associated with toric morphisms to a singular base, where some rays
are not included in the base.  Furthermore, there can be cases where a
triangulation is such that there is no toric morphism to any base
associated with the projection of $\nabla_2$ to the origin, and  there
are cases where there is no triangulation at all satisfying the usual
(fine, regular, star) conditions that is compatible with such a toric
morphism.  Nonetheless, we can demonstrate that whenever there is a
reflexive subpolytope $\nabla_2 \subset \nabla$ there is always a
triangulation of $\nabla$ giving a toric variety with a fan $\Sigma$
compatible with a toric morphism $\pi:\Sigma \rightarrow
\Sigma_{B_s}$, where $B_s$ is the (generally singular) toric base
defined by taking only the primitive rays associated with the
projection of certain vertices of the polytope $\nabla$ under the map
$\pi: \nabla_2 \rightarrow 0$ (in many cases in practice the base can
be taken to be a refinement $B'$ of $B_s$).  This triangulation can be
constructed by starting with a triangulation of the 4D polytope
defined by $\nabla_2$ and a set of points associated with (necessarily
primitive) vertices $u_i$ of $\nabla$ that project to the vertices of
$\pi (\nabla)$, using the natural orderings of rays in $\nabla_2$ and
$\pi (\nabla)$.  The resulting toric variety is generally highly
singular, and the fan can then be refined by including the remaining
lattice points of $\nabla$ in an arbitrary refinement of the original
singular triangulation; this final triangulation respects the toric
morphism to $B_s$ (or a refinement $B'$ thereof), and gives a genus
one or elliptic fibration of the resulting toric variety.  The fan
constructed in this way, however, generally does not give a
triangulation of the full polytope $\nabla$ that satisfies the star
condition (though it is a star triangulation of a non-convex polytope
contained within $\nabla$), so that the resulting Calabi-Yau phase
does not directly fit into the Batyrev framework and requires the more
general structure of ``vex'' polytopes for toric varieties that are
not weak Fano developed in \cite{Berglund-Hubsch}. The need to include
ambient toric varieties where the anticanonical class has a base locus
is related to the presence of ``non-Higgsable clusters''
\cite{clusters}, associated with rigid subvarieties of the Calabi-Yau
hypersurface that are present everywhere in the moduli space of that
phase.  The presence of a triangulation of this type guarantees a
Calabi-Yau phase with an elliptic or genus one fibration, although
particularly in higher dimensions there can be singularities without a
Calabi-Yau resolution \cite{Klemm-lry, MC}. The issue of regularity,
corresponding to positivity in the K\"ahler cone, must also be
addressed.  We leave a more detailed explication of these issues to
future work; the upshot of this discursion is that the presence of a
reflexive 2D subpolytope $\nabla_2 \subset \nabla$ is generally
associated with at least one triangulation giving an elliptic or genus
one fibration of a corresponding toric variety.

We should also emphasize that while the presence of a reflexive 2D
subpolytope seems in general to be sufficient for a genus one or
elliptic fiber of an associated Calabi-Yau threefold, this condition
is not necessary.  In studying subpolytopes of $\nabla$, we are only
finding the ``obvious'' fibers that are encoded in a natural way in
the toric structure.  There can also be fibrations that are not
represented torically, which could be analyzed by a more complete
treatment of the structure of the Calabi-Yau threefold using the full
set of
triple intersection numbers and K\"ahler cone information, as was done
for CICY threefolds in \cite{aggl-3}.

\section{Results} 
\label{sec:results}

\subsection{Fiber analysis of all reflexive 4D polytopes}

Extending the work initiated in \cite{Huang-Taylor}, we have checked
all 473.8 million reflexive  4D polytopes in the Kreuzer-Skarke
database \cite{database} for the presence of at least one reflexive 2D
subpolytope passing through the origin.  We find that all but 29,223
(i.e.\ 99.994\%)
of the 4D reflexive polytopes have such a fiber.

\begin{figure}
\begin{center}
\includegraphics[width=10cm]{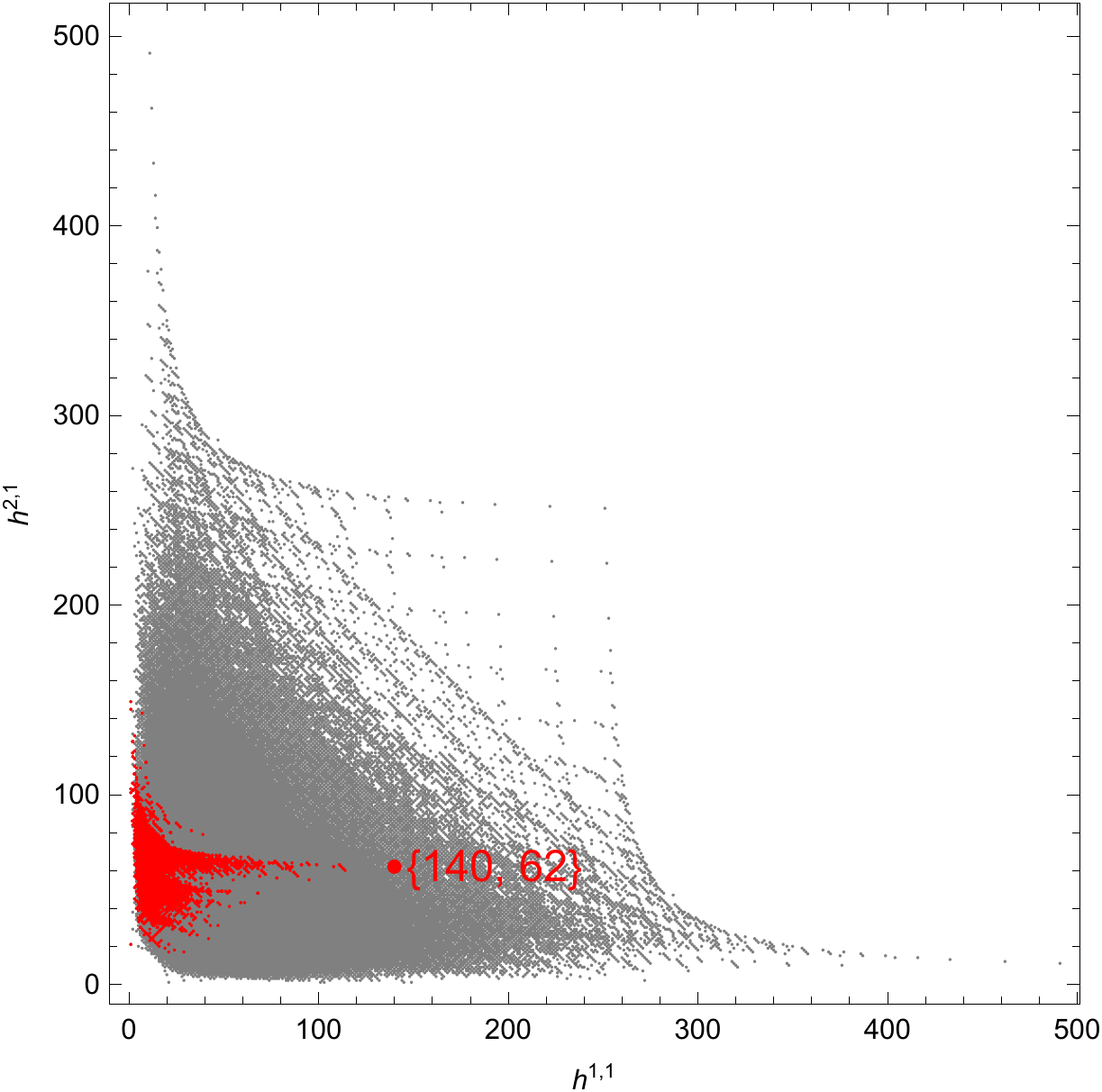}
\end{center}
\caption[x]{\footnotesize The Hodge numbers of the Calabi-Yau
  threefolds associated with 4D reflexive polytopes lacking a 2D
  reflexive subpolytope (red), plotted over the Hodge numbers of the
  full set of toric hypersurface Calabi-Yau threefolds (gray).}
\label{f:29k}
\end{figure}

The 1,395 distinct Hodge numbers of the 29,223 cases without reflexive 2D
subpolytopes are shown in red in Figure~\ref{f:29k}.
The largest $h^{1,1}$ for a  Calabi-Yau threefold associated with a
non-fibered 4D polytope  comes from the case with Hodge numbers (140,
62), as previously identified in \cite{Huang-Taylor}.
The next-largest values of $h^{1,1}$ come from polytopes associated
with Calabi-Yau threefolds having Hodge numbers
\begin{equation}
 (114, 60), (113, 61) \times 2, (112, 62), (111, 63) \times 2,
  (108, 62) \times 2, (99, 63) \times 2, \ldots
\end{equation}
It is very interesting that the polytopes corresponding with large
$h^{1,1}$ that do not have 2D reflexive subpolytopes have Hodge
numbers $h^{2,1}$ in the narrow range 60---63; the first exception as
$h^{1,1}$ decreases has the Hodge numbers (95, 55), and for all
$h^{1,1} > 61$ the non-fibered cases have $h^{2,1}$ in the range
48---65.  It would be interesting to understand better whether this
family of polytopes has some common structure associated with the lack
of an elliptic or genus one fibration for the corresponding toric
varieties.
Indeed, there are similarities between some of these cases: in
particular, for example, the polytope giving Hodge numbers (114, 60) has only  a
slight difference in one vertex from one of the polytopes giving Hodge
numbers (113, 61); we leave a more detailed analysis of common
structure in these polytopes for further work, however.

\begin{figure}
\begin{center}
\includegraphics[width=14cm]{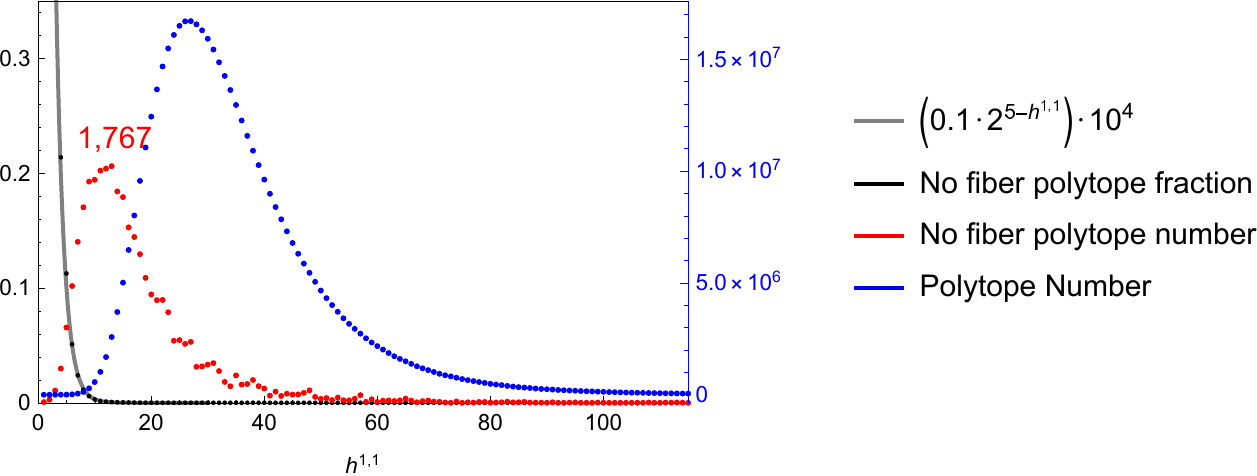}
\end{center}
\caption[x]{\footnotesize  The distribution of 4D reflexive polytopes
  without a 2D reflexive subpolytope fiber (red) compared to the total
number of 4D reflexive polytopes (blue) as a function of Hodge number
$h^{1, 1}$, and the fraction without fibers (gray).
Note that the vertical axis is normalized differently for each of the
three curves.}
\label{f:graph}
\end{figure}

The distribution of polytopes without a 2D reflexive subpolytope is
graphed in Figure~\ref{f:graph}, and compared to the total number of
polytopes.  While the total number of reflexive 4D polytopes at a
given value of $h^{1,1}$ peaks at $h^{1,1} = 27$ (with 16.7 million reflexive 4D
polytopes at this value of $h^{1,1}$), the number without a
reflexive 2D subpolytope peaks at $h^{1,1} = 13$ (with 1,767 cases).

We have made available the results of our analysis at  \cite{data},
where the list of 29,223 4D reflexive polytopes without a 2D reflexive
subpolytope can be found in the format of \cite{database}.

\subsection{Asymptotics at small $h^{1,1}$}

\begin{figure}
\begin{center}
\includegraphics[width=10cm]{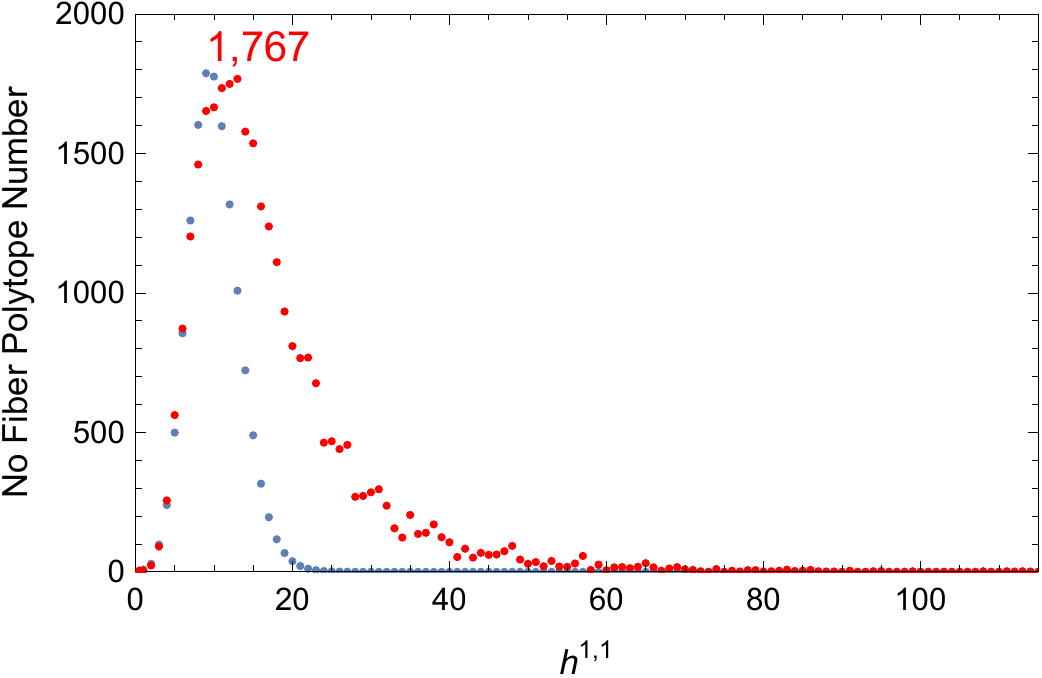}
\end{center}
\caption[x]{\footnotesize Illustration of the ``fat tail'' in the
  number of 4D reflexive polytopes without 2D reflexive toric fibers
  (red), compared to exponentially suppressed prediction (gray) based
  on heuristic analysis of elliptic fibrations for Calabi-Yau
  threefolds from \cite{Huang-Taylor}.}
\label{f:comparison}
\end{figure}

In \cite{Huang-Taylor}, a heuristic argument was given for why the
fraction of Calabi-Yau threefolds without an elliptic or genus one
fiber may decrease exponentially, and it was noted that at small
values of $h^{1,1}$ the empirically computed value of this fraction
from the KS database decreases roughly as $2^{- h^{1,1}}$. The
heuristic argument is based on the assumption that the triple
intersection numbers of the Calabi-Yau are essentially random and that
the K\"ahler cone is a generalized quadrant defined by conditions $x_i
\geq 0$ on the coefficients of the K\"ahler class in a natural basis
of effective divisors.  With the full set of data, we see that while
this is not a bad approximation at small $h^{1,1}$, the tail of the
distribution is considerably fatter; indeed, as pointed out in
\cite{Huang-Taylor}, the distribution must drop off less rapidly than
$2^{- h^{1,1}}$ for the large $h^{1,1}$ examples such as (140, 62) to
lack fibers.  The tail of the observed distribution is compared to the
prediction based on the asymptotic estimate $c \cdot 2^{- h^{1,1}}$
fit to the small values of $h^{1,1}$ in Figure~\ref{f:comparison}.

It is an interesting question for further work to determine why the
 observed distribution drops off less quickly than the
exponential estimate.  One possibility is that the toric structure is
simply missing the elliptic/genus one fibration structure in many of
the cases with larger values of $h^{1, 1}$.  While the
presence of a reflexive 2D subpolytope is an indicator for an elliptic
or genus one fibration for at least one phase of the associated
Calabi-Yau threefold, such a subpolytope is not a necessary condition
for a fibration.  A fibration  of an associated Calabi-Yau threefold
may occur in a way that is simply not captured by the toric geometry
of the ambient space into which the Calabi-Yau is embedded as a
hypersurface.  A similar situation was encountered in the analysis of
the CICY Calabi-Yau threefolds in \cite{aggl-3}; while many of the CICYs
have ``obvious'' fibrations that can be seen in the structure of the
defining matrices, there are also non-obvious fibrations that can be
identified by a more careful analysis of the triple intersection
numbers and K\"ahler cone of the Calabi-Yau.  One might similarly
expect that many non-toric fibrations may be present for the cases for
which we have not identified a reflexive 2D subpolytope; this
would be
interesting to investigate further.  Another
possibility is that there is an error coming from the assumption that the
K\"ahler cone can be approximated as a generalized quadrant.  Indeed,
a recent analysis \cite{narrow-Kaehler} suggests that at large $h^{1,1}$ the K\"ahler cone
becomes progressively more narrow.  This may lead to a situation that
makes it easier to avoid having an elliptic or genus one fibration as
$h^{1,1}$ increases than the naive argument leading to the exponential
decrease would suggest.  Nonetheless, even though the tail of the
distribution is a bit fatter than the naive exponential argument
suggested, the drop-off is still relatively quick and the  total number of
reflexive 4D polytopes without a reflexive 2D subpolytope is really
quite small.

\subsection{Mirror symmetry}

One interesting application of this analysis may be to mirror
symmetry.  It was pointed out in \cite{HT-mirror} that in many cases
where a polytope and its dual both admit a reflexive 2D subpolytope
the mirror symmetry between the associated Calabi-Yau threefolds
essentially factorizes  between the fiber and the base.  Since we have
found fibers for most reflexive 4D polytopes, it suggests that this
factorization may be relevant for most Calabi-Yau threefolds.

One interesting class of cases for further study are those where one
side of a mirror pair is elliptic or genus one fibered and the other
is not.  For example, the mirror quintic with Hodge numbers (101, 1)
has a fibration, while the quintic (like all Calabi-Yau threefolds
with $h^{1, 1} = 1$) cannot have an elliptic or genus one fiber.  One
might hope to connect non-fibered threefolds like the quintic with the
large connected set of elliptic Calabi-Yau threefolds through a
transition, which may be related to how the mirror quintic is
connected to other elliptic Calabi-Yau threefolds.

It is also interesting to consider cases where neither the polytope
nor its mirror admit a manifest toric
fibration.  A systematic check shows that there
are only 24 such polytopes $\nabla$ in the KS database that do not have a 2D
reflexive subpolytope either for $\nabla$ or its dual $\Delta$.  These
24 polytopes have the Hodge numbers
\begin{equation}
(43, 59), (44, 44), (63, 67), (64, 66), (65, 65) \,.
\end{equation}
It would be interesting to understand better whether there is some
common structure underlying these polytopes and their associated
Calabi-Yau threefolds, and whether they completely lack elliptic/genus one fibers or such fibers are simply not visible torically.

\section{Discussion}
\label{sec:discussion}

We have carried out a systematic analysis of the full database of
473.8 million reflexive 4D polytopes and found that fewer than 30,000
of these lack a 2D reflexive subpolytope.  It seems that the 99.994\%
of polytopes with a 2D reflexive fiber correspond to Calabi-Yau
threefolds that are birationally equivalent to a Calabi-Yau threefold
phase that has a genus one or elliptic fibration, though some
technical aspects of the triangulations associated with the fibered
phase of some polytopes remain to be resolved.
This gives strong evidence supporting the hypothesis that most
Calabi-Yau threefolds are elliptic or genus one fibered, up to
birational equivalence; this in turn, if true, would indicate that the
number of birational equivalence classes of Calabi-Yau threefolds is
finite.

The apparent predominance of elliptic and genus one fibrations among
Calabi-Yau threefolds has a number of interesting implications for
physics. In particular, this has implications for the role of F-theory
in understanding the global set of string compactifications.  For 6D
supergravity theories, F-theory seems to give a good global
description of the space of possible string compactifications, and
essentially all known string compactifications to ${\cal N} = 1$ 6D
supergravity theories have a dual description in F-theory.  This is
not true in 4D, however; for example, heterotic compactification on
a Calabi-Yau threefold only has a known F-theory dual description when
the Calabi-Yau threefold is elliptic or genus one fibered (see
e.g. \cite{FMW}).  If, however, indeed most Calabi-Yau threefolds have
such a fibration, then most heterotic compactifications have an
F-theory dual description. Thus, the results of this paper contribute
evidence to the hypothesis that F-theory gives a very general
representative sample of ${\cal N} = 1$ 4D string theory vacua.

The analysis here of toric hypersurface Calabi-Yau threefolds and the
analysis in \cite{aggl-2, aggl-3} of CICY threefolds give similar
results for the overwhelming dominance of elliptic and genus one
fibered Calabi-Yau threefolds for these large families.  It would be
interesting to extend this kind of analysis to other known Calabi-Yau
threefolds that do not fit in these two families.  A variety of other
constructions can give other Calabi-Yau threefolds, particularly at
small Hodge numbers (see \cite{Davies} for a review), and it would be
interesting to analyze the fibration structure of these other
constructions.  It would also be a good check on these results to
consider the fibration structure of more general complete
intersections in toric varieties; the fibers that may arise in such
constructions were systematically analyzed in \cite{Braun:2014qka}.

The analysis of this paper may be helpful in various systematic studies of
Calabi-Yau threefolds and their properties that are relevant for physics, since there are many
questions that are easier to answer for elliptic and genus one fibered
Calabi-Yau threefolds than for general Calabi-Yau threefolds (see
e.g.\ \cite{hkk}). It would also be interesting to understand better the
relatively small class of 4D reflexive polytopes without a reflexive
2D subpolytope; these correspond to Calabi-Yau threefolds without an
(obvious) elliptic or genus one fiber, which seem in light of this
analysis to form a rather special subset of the set of known
Calabi-Yau threefolds.

One can ask similar questions about elliptic and genus one fibration
structure for Calabi-Yau fourfolds.  In \cite{Gray-hl} it was shown
that the fraction of CICY Calabi-Yau fourfolds that has an obvious
elliptic or genus one fibration (99.95\%) is even larger than the
fraction of CICY Calabi-Yau threefolds with this property (99.3\%).
There is no complete analysis of reflexive 5D polytopes, though there
are some partial results in this direction \cite{ss} and some analysis
of the fibration structure of these polytopes was carried out in
\cite{Rohsiepe}.  In fact, direct construction of toric threefold
bases that support elliptic and genus one toric hypersurface
Calabi-Yau fourfolds shows that the number of such bases alone is
extraordinarily large, on the order of $10^{3000}$ \cite{MC,
  Halverson-ls, Wang-WT-MC-2}.  Thus, the number of elliptic
Calabi-Yau fourfolds that can be directly constructed in this way
overwhelmingly exceeds any other class of known constructions for
Calabi-Yau fourfolds.  It is thus the case that the vast majority of
known constructions of Calabi-Yau fourfolds are automatically elliptic
or genus one fibered.

\acknowledgments{
We would like to thank Lara Anderson, Anuj Apte, Per Berglund,
Antonella Grassi, James Gray, Jim Halverson, Cody Long, Liam
McAllister, Brent Nelson, Paul Oehlmann, Sakura Schafer-Nameki, Yi-Nan
Wang and S.\-T.\ Yau for helpful discussions.  This material is based
upon work supported by the U.S.\ Department of Energy, Office of
Science, Office of High Energy Physics under grant Contract Number
DE-SC00012567.  WT would like to thank the high-energy theory physics
groups at UCSD and UCSC for hospitality during the time when most of
this work was carried out.  }

\newpage
\appendix


\begin{thebibliography}{99}


\bibitem{chsw}
  P.~Candelas, G.~T.~Horowitz, A.~Strominger and E.~Witten,
  ``Vacuum Configurations for Superstrings,''  Nucl.\ Phys.\ B {\bf 258}, 46 (1985).  

\bibitem{Yau-66} 
S.-T.\  Yau, ``Open problems in geometry,''
Proc.\ Symp.\ Pure Math {\bf  54}, 1, 1-28 (1993).

\bibitem{Gross}
M.\ 
Gross, 
``A finiteness theorem for elliptic Calabi--Yau threefolds,''
Duke Math.\ Jour.\  {\bf  74}, 271 (1994). 

\bibitem{Vafa-F-theory}
  C.~Vafa,
  ``Evidence for F-Theory,''
  Nucl.\ Phys.\  B {\bf 469}, 403 (1996)
  {\tt arXiv:hep-th/9602022}.

\bibitem{Morrison-Vafa-I}
  D.~R.~Morrison and C.~Vafa,
  ``Compactifications of F-Theory on Calabi--Yau Threefolds -- I,''
  Nucl.\ Phys.\  B {\bf 473}, 74 (1996)
  {\tt arXiv:hep-th/9602114};
    

\bibitem{Morrison-Vafa-II}    
  D.~R.~Morrison and C.~Vafa,
  ``Compactifications of F-Theory on Calabi--Yau Threefolds -- II,''
  Nucl.\ Phys.\  B {\bf 476}, 437 (1996)
  {\tt arXiv:hep-th/9603161}.

\bibitem{Grassi}
A.~Grassi, ``On minimal models of elliptic threefolds,'' Math. Ann. {\bf 290}
  (1991) 287--301.

\bibitem{KMT-2}
  V.~Kumar, D.~R.~Morrison and W.~Taylor,
  ``Global aspects of the space of 6D ${\cal N} = 1$ supergravities,''
  JHEP {\bf 1011}, 118 (2010)
  {\tt arXiv:1008.1062 [hep-th]}.

\bibitem{toric} 
  D.~R.~Morrison and W.~Taylor,
  ``Toric bases for 6D F-theory models,''
  Fortsch.\ Phys.\  {\bf 60}, 1187 (2012)
  {\tt arXiv:1204.0283 [hep-th]}.
  

\bibitem{Hodge} 
  W.~Taylor,
  ``On the Hodge structure of elliptically fibered Calabi-Yau threefolds,''
JHEP {\bf 1208}, 032 (2012)
{\tt arXiv:1205.0952 [hep-th]}.

\bibitem{Wang-WT}
W.~Taylor, Y.~N.~Wang,
``Non-toric bases for elliptic Calabi-Yau Threefolds and 6D F-theory Vacua,''
{\tt arXiv:1504.07689}. 

\bibitem{Johnson-WT}
S.~Johnson, W.~Taylor,
``Calabi-Yau Threefolds with Large $h^{2,1}$,''
JHEP {\bf 1410}, 23 (2014),
{\tt arXiv:1406.0514}.

\bibitem{Johnson-WT-2} 
  S.~B.~Johnson and W.~Taylor,
  ``Enhanced gauge symmetry in 6D F-theory models and tuned elliptic Calabi-Yau threefolds,''
Fortsch.\ Phys.\  {\bf 64}, 581 (2016)
{\tt arXiv:1605.08052 [hep-th]}.

\bibitem{cdls}
  P.~Candelas, A.\ M.\ Dale, C.\ A.\ Lutken and R.\ Schimmrigk,
  `` Complete Intersection Calabi-Yau Manifolds,''  Nucl.\ Phys.\ B
  {\bf 298}, 493  (1988).

\bibitem{Anderson-aggl} 
  L.~B.~Anderson, F.~Apruzzi, X.~Gao, J.~Gray and S.~J.~Lee,
  ``A New Construction of Calabi-Yau Manifolds: Generalized CICYs,''
Nucl.\ Phys.\ B {\bf 906}, 441 (2016)
{\tt arXiv:1507.03235 [hep-th]}.

\bibitem{KS-classification} 
  M.~Kreuzer and H.~Skarke,
  ``Complete classification of reflexive polyhedra in four-dimensions,''
  Adv.\ Theor.\ Math.\ Phys.\  {\bf 4}, 1209 (2002)
  {\tt hep-th/0002240}.
  

\bibitem{aggl-2} 
  L.~B.~Anderson, X.~Gao, J.~Gray and S.~J.~Lee,
  ``Multiple Fibrations in Calabi-Yau Geometry and String Dualities,''
JHEP {\bf 1610}, 105 (2016)
{\tt arXiv:1608.07555 [hep-th]}.

\bibitem{aggl-3} 
  L.~B.~Anderson, X.~Gao, J.~Gray and S.~J.~Lee,
  ``Fibrations in CICY Threefolds,''
JHEP {\bf 1710}, 077 (2017)
{\tt arXiv:1708.07907 [hep-th]}.

\bibitem{agh-non-simply} 
  L.~B.~Anderson, J.~Gray and B.~Hammack,
  ``Fibrations in Non-simply Connected Calabi-Yau Quotients,''
  {\tt arXiv:1805.05497 [hep-th]} .

\bibitem{Candelas-cs} 
  P.~Candelas, A.~Constantin and H.~Skarke,
  ``An Abundance of K3 Fibrations from Polyhedra with Interchangeable
  Parts,''  Commun.\  Math.\  Phys.\  {\bf 324}, 937 (2013)
{\tt arXiv:1207.4792 [hep-th]}.  

\bibitem{database} 
M. Kreuzer and H. Skarke, 
\url{http://hep.itp.tuwien.ac.at/~kreuzer/CY.html}.

\bibitem{Huang-Taylor-long}
  Y.~C.~Huang and W.~Taylor,
``Comparing elliptic and toric hypersurface Calabi-Yau threefolds at
  large Hodge numbers,''
    JHEP {\bf 1902}, 087 (2019)
{\tt  arXiv:1805.05907 [hep-th]}.

\bibitem{Huang-Taylor} 
  Y.~C.~Huang and W.~Taylor,
  ``On the prevalence of elliptic and genus one fibrations among toric hypersurface Calabi-Yau threefolds,''
  JHEP {\bf 1903}, 014 (2019)
{\tt arXiv:1809.05160 [hep-th]}.

\bibitem{Batyrev}
  V.\ Batyrev,
  ``Variations of the  Mixed Hodge Structure of Affine
  Hypersurfaces in Algebraic Tori,''
Duke Math.\ Journ.\ {\bf 69}, 349 (1993).

\bibitem{Kreuzer-Skarke-fibers} 
  M.~Kreuzer and H.~Skarke,
  ``Calabi-Yau four folds and toric fibrations,''
  J.\ Geom.\ Phys.\  {\bf 26}, 272 (1998)
  {\tt hep-th/9701175}.

\bibitem{Rohsiepe} 
  F.~Rohsiepe,
  ``Fibration structures in toric Calabi-Yau fourfolds,''
  {\tt arXiv:hep-th/0502138.}

\bibitem{Braun} 
  V.~Braun,
  ``Toric Elliptic Fibrations and F-Theory Compactifications,''
  JHEP {\bf 1301}, 016 (2013)
  {\tt arXiv:1110.4883 [hep-th]}.
  

\bibitem{Bouchard-Skarke} 
  V.~Bouchard and H.~Skarke,
  ``Affine Kac-Moody algebras, CHL strings and the classification of tops,''
Adv.\ Theor.\ Math.\ Phys.\  {\bf 7}, no. 2, 205 (2003)
{\tt hep-th/0303218}.

\bibitem{BGK-geometric} 
  V.~Braun, T.~W.~Grimm and J.~Keitel,
  ``Geometric Engineering in Toric F-Theory and GUTs with U(1) Gauge Factors,''
JHEP {\bf 1312}, 069 (2013)
{\tt arXiv:1306.0577 [hep-th]}.

\bibitem{Klevers-16} 
  D.~Klevers, D.~K.~Mayorga Pena, P.~K.~Oehlmann, H.~Piragua and J.~Reuter,
  ``F-Theory on all Toric Hypersurface Fibrations and its Higgs Branches,''
JHEP {\bf 1501}, 142 (2015)
{\tt arXiv:1408.4808 [hep-th]}.

\bibitem{HT-mirror} 
  Y.~C.~Huang and W.~Taylor,
  ``Mirror symmetry and elliptic Calabi-Yau manifolds,''
  JHEP {\bf 1904}, 083 (2019)
{\tt arXiv:1811.04947 [hep-th]}.

\bibitem{data}
See 
{\tt http://ctp.lns.mit.edu/wati/data.html}

\bibitem{tops} 
  P.~Candelas and A.~Font,
  ``Duality between the webs of heterotic and type II vacua,''
  Nucl.\ Phys.\ B {\bf 511}, 295 (1998)
  {\tt hep-th/9603170}.
  

\bibitem{Per-forthcoming}
P.\ Berglund, Y.\ C.\ Huang, W.\ Taylor and Y.-N.\ Wang, {\it to appear};

\bibitem{cj-forthcoming}
J.\ Halverson, C.\ Long and W.\ Taylor, {\it to appear}.

\bibitem{Candelas-nf} 
  P.~Candelas, D.~E.~Diaconescu, B.~Florea, D.~R.~Morrison and G.~Rajesh,
  ``Codimension three bundle singularities in F theory,''
  JHEP {\bf 0206}, 014 (2002)  {\tt hep-th/0009228}.

\bibitem{Lawrie-Schafer-Nameki} 
  C.~Lawrie and S.~Sch\"afer-Nameki,
  ``The Tate Form on Steroids: Resolution and Higher Codimension Fibers,''
  JHEP {\bf 1304}, 061 (2013)
  {\tt arXiv:1212.2949 [hep-th]}.

\bibitem{Borchmann:2013jwa} 
  J.~Borchmann, C.~Mayrhofer, E.~Palti and T.~Weigand,
  ``Elliptic fibrations for $SU(5)\times U(1)\times U(1)$ F-theory vacua,''
  Phys.\ Rev.\ D {\bf 88}, no. 4, 046005 (2013)
  {\tt arXiv:1303.5054 [hep-th]}.

\bibitem{Cvetic:2013uta} 
  M.~Cveti\v{c}, A.~Grassi, D.~Klevers and H.~Piragua,
  ``Chiral Four-Dimensional F-Theory Compactifications With SU(5) and Multiple U(1)-Factors,''
  JHEP {\bf 1404}, 010 (2014)
  {\tt arXiv:1306.3987 [hep-th]}.

 

\bibitem{dor-nf} 
  M.~Dierigl, P.~K.~Oehlmann and F.~Ruehle,
  ``Global Tensor-Matter Transitions in F-Theory,''
  Fortsch.\ Phys.\  {\bf 66}, no. 7, 1800037 (2018)
  {\tt arXiv:1804.07386 [hep-th]}.


\bibitem{Achmed-Zade:2018idx} 
  I.~Achmed-Zade, I.~García-Etxebarria and C.~Mayrhofer,
  ``A note on non-flat points in the SU(5) × U(1)$_{PQ}$ F-theory model,''
  JHEP {\bf 1905}, 013 (2019)
  {\tt arXiv:1806.05612 [hep-th]}.

\bibitem{Tian-Wang} 
  J.~Tian and Y.~N.~Wang,
  ``E-string spectrum and typical F-theory geometry,''
  {\tt arXiv:1811.02837 [hep-th]}.

\bibitem{alm-nf} 
  F.~Apruzzi, L.~Lin and C.~Mayrhofer,
  ``Phases of 5d SCFTs from M-/F-theory on Non-Flat Fibrations,''
  JHEP {\bf 1905}, 187 (2019)
  {\tt arXiv:1811.12400 [hep-th]}.

\bibitem{Apruzzi:2019opn} 
  F.~Apruzzi, C.~Lawrie, L.~Lin, S.~Schafer-Nameki and Y.~N.~Wang,
  ``Fibers add Flavor, Part I: Classification of 5d SCFTs, Flavor Symmetries and BPS States,''
  {\tt arXiv:1907.05404 [hep-th]}.


\bibitem{Berglund-Hubsch} 
  P.~Berglund and T.~Hubsch,
  ``A Generalized Construction of Calabi-Yau Models and Mirror Symmetry,''
  SciPost Phys.\  {\bf 4}, no. 2, 009 (2018)
{\tt arXiv:1611.10300 [hep-th]}.

\bibitem{clusters} 
  D.~R.~Morrison and W.~Taylor,
  ``Classifying bases for 6D F-theory models,''
Central Eur.\ J.\ Phys.\  {\bf 10}, 1072 (2012),
{\tt arXiv:1201.1943 [hep-th]}.

\bibitem{Klemm-lry} 
  A.~Klemm, B.~Lian, S.~S.~Roan and S.~T.~Yau,
  ``Calabi-Yau fourfolds for M theory and F theory compactifications,''
Nucl.\ Phys.\ B {\bf 518}, 515 (1998)
{\tt hep-th/9701023}.

\bibitem{MC} 
  W.~Taylor and Y.~N.~Wang,
  ``A Monte Carlo exploration of threefold base geometries for 4d F-theory vacua,''
  JHEP {\bf 1601}, 137 (2016),
 {\tt  arXiv:1510.04978 [hep-th]}.

\bibitem{narrow-Kaehler} 
  M.~Demirtas, C.~Long, L.~McAllister and M.~Stillman,
  ``The Kreuzer-Skarke Axiverse,''
  {\tt arXiv:1808.01282 [hep-th]}.

\bibitem{FMW} 
  R.~Friedman, J.~Morgan and E.~Witten,
  ``Vector bundles and F theory,''
Commun.\ Math.\ Phys.\  {\bf 187}, 679 (1997)
{\tt hep-th/9701162}.


\bibitem{Davies} 
  R.~Davies,
  ``The Expanding Zoo of Calabi-Yau Threefolds,''  Adv.\ High Energy
  Phys.\  {\bf 2011}, 901898 (2011)  {\tt arXiv:1103.3156 [hep-th]}.

\bibitem{Braun:2014qka} 
  V.~Braun, T.~W.~Grimm and J.~Keitel,
  ``Complete Intersection Fibers in F-Theory,''
JHEP {\bf 1503}, 125 (2015)
{\tt arXiv:1411.2615 [hep-th]}.


\bibitem{hkk} 
  M.~x.~Huang, S.~Katz and A.~Klemm,
  ``Topological String on elliptic CY 3-folds and the ring of Jacobi forms,''
  JHEP {\bf 1510}, 125 (2015)
  {\tt arXiv:1501.04891 [hep-th]}.

\bibitem{Gray-hl} 
  J.~Gray, A.~S.~Haupt and A.~Lukas,
  ``Topological Invariants and Fibration Structure of Complete
  Intersection Calabi-Yau Four-Folds,''
JHEP {\bf 1409}, 093 (2014)
{\tt arXiv:1405.2073 [hep-th]}.

\bibitem{ss} 
  F.~Schoeller and H.~Skarke,
  ``All Weight Systems for Calabi-Yau Fourfolds from Reflexive Polyhedra,''
  {\tt arXiv:1808.02422 [hep-th]}.


\bibitem{Halverson-ls} 
  J.~Halverson, C.~Long and B.~Sung,
  ``Algorithmic universality in F-theory compactifications,''
  Phys.\ Rev.\ D {\bf 96}, no. 12, 126006 (2017)
  {\tt arXiv:1706.02299 [hep-th]}.

\bibitem{Wang-WT-MC-2} 
  W.~Taylor and Y.~N.~Wang,
  ``Scanning the skeleton of the 4D F-theory landscape,''
  JHEP {\bf 1801}, 111 (2018)
  {\tt arXiv:1710.11235 [hep-th]]}.

\end{thebibliography}
\end{document}